\documentstyle[12pt]{article}
\begin{document}
\begin{titlepage}
\renewcommand{\thefootnote}{\fnsymbol{footnote}}
\makebox[2cm]{}\\[-1in]
\begin{flushright}
\begin{tabular}{l}
CAMK 95-288\\
hep-ph/9502271
\end{tabular}
\end{flushright}
\vskip0.4cm
\begin{center}
{\Large\bf Phenomenological determination of polarized quark
distributions in the nucleon }\footnote {Work supported in part
by the KBN Grant 2-P302-143-06}

\vspace{2cm}

Jan Bartelski$^1$ and Stanis\l aw Tatur$^2$

\vspace{1.5cm}

$^1${\em Institute of Theoretical Physics, Warsaw University,\\
Ho$\dot{z}$a 69, 00-681 Warsaw, Poland.}\\[0.5cm]
$^2${\em N. Copernicus Astronomical Center,
        Polish Academy of Science,\\
        ul. Bartycka 18, 00-716 Warsaw, Poland.  }\\[0.5cm]

\vspace{1cm}


\vspace{1cm}

{\bf Abstract:\\[5pt]}
\parbox[t]{\textwidth}{ We present a fit to spin asymmetries which gives
polarized quark distributions. These functions are
closely related to the ones given by the newest Martin, Roberts
and Stirling fit for unpolarized structure functions. The integrals of
polarized distributions are discussed and compared with the
corresponding quantities obtained from neutron and hyperon
$\beta$-decay data. We use the combination of proton and
neutron spin asymmetries in order to determine the
coefficients of our polarized quark distributions. Our fit shows
that phenomenologically there is no need for taking gluonic
degrees of freedom into account.  }

\vspace{1cm}
\end{center}
\end{titlepage}
\newpage
The new data taken in experiments at CERN \cite{d,p} and SLAC
\cite{n} renewed the interest in the spin structure of the nucleon.
We have now at our disposal relatively precise CERN data
for proton spin asymmetries from Spin Muon Collaboration (SMC) \cite{p}
and very recent E143 experiment results from SLAC \cite{e143}. We have
also the asymmetries measured from $^{3}He$  in E142 experiment at SLAC
\cite{n}
whereas the data from
deuterium target come from SMC \cite{d}. Together with an old SLAC
\cite{pold} and EMC \cite{emc} data for proton one has considerable amount
of information which can be used to study nucleon spin structure and in
particular to determine the polarized quark distributions.

The unpolarized quark distributions in the nucleon are known
due to several fits [7-10]. Martin, Roberts
and Stirling (MRS) \cite{mrs} gave a complete fit using an
existing experimental data in order
to determine parton (i.e. quark and gluon)
distributions. Quite recently an improved version of such
densities has been presented by them \cite{mrs1}.

Our preliminary discussion how to determine the polarized
quark distributions using the older version $D_{-}^{'}$ of the
MRS fit was given in \cite{bt}.

In this paper we would like to get polarized quark parton distributions
starting from the unpolarized ones and using existing data for proton, neutron
(made on $^{3}He$) and deuteron spin asymmetries. We will consider two
different models of possible $x$ behaviour (more and less singular at $x
\rightarrow 0$) of the sea contribution. The calculated values of octet
axial-vector couplings $a_{3}$ and $a_{8}$ obtained from the fit are compared
with the experimental values gotten from nucleon and hyperon $\beta$-decays
modified for QCD corrections. These quantities are used to
differentiate  between various fits
(together with their $\chi^{2}$ values) which were made for different
combinations of the spin
asymmetry data (e.g. proton+neutron versus proton+deuteron data). Putting the
additional restriction for $a_{8}$ (in order to stabilize the
fits) we determine
polarized quark distributions from the combination of {\em
proton} and {\em neutron} data.
We have also tried to include polarized gluons contributing in the way
proposed in Ref.{\cite{ross}. It does not lead to any substantial improvement
($\chi^{2}$ per degree of freedom is worse) in the fit. In addition the
sign of the gluon
contribution is opposite to the one which is expected
theoretically. The inclusion
of newest SLAC data for proton asymmetry from the E143 experiment is also
discussed.

Let us start with the formulas for unpolarized quark parton distributions
(at $ Q^{2}=4\, {\rm GeV^{2}}$) given by Martin, Roberts and Stirling
\cite{mrs1}. We shall consider their fit called
MRS(A) with rather singular behavior of
sea distribution at small $x$ values (which, however agrees with the
results from HERA \cite{hera}). We have for the valence quarks distributions:
\begin{eqnarray}
u_{v}(x)&=&1.996 x^{-0.462}(1-x)^{3.96}(1-0.39\sqrt{x}+5.13x),
\nonumber \\
d_{v}(x)&=&0.296 x^{-0.670}(1-x)^{4.71}(1+5.03\sqrt{x}+5.56x) ,
\end{eqnarray}

\noindent and for the sea ones:
\begin{eqnarray}
2\bar{u} (x)&=&0.392M(x)-\delta (x), \nonumber \\
2\bar{d} (x)&=&0.392M(x)+\delta (x), \nonumber \\
2\bar{s} (x)&=&0.196M(x), \\
2\bar{c} (x)&=&0.020M(x). \nonumber
\end{eqnarray}

\noindent In eq.(2) one has for singlet:
\begin{equation}
M(x)=0.411x^{-1.3}(1-x)^{9.27}(1-1.15\sqrt{x}+15.6x),
\end{equation}

\noindent and isovector part:
\begin{equation}
\delta (x)=0.099x^{-0.462}(1-x)^{9.27}(1+25.0x).
\end{equation}

\noindent The unpolarized gluon distribution is given by:
\begin{equation}
G(x)=0.775x^{-1.3}(1-x)^{5.3}(1+5.2x).
\end{equation}

In analogy to the unpolarized case we assume that the polarized
quark distributions are of the form:
$x^{\alpha}(1-x)^{\beta}P_{2}( \sqrt{x})$, where
$P_{2}(\sqrt{x})$ is a second order polynomial in $\sqrt{x}$ and
the asymptotic behavior for $x$$\rightarrow$$0$ and
$x$$\rightarrow$$1$ (i.e. the values of $\alpha$ and $\beta$) are the same as
in
unpolarized case.
The unpolarized parton distribution is the sum of
helicity distributions along and opposite to parent nucleon
helicity whereas the polarized one is the difference of
such functions. Hence, our idea is just to split the numerical constants
(coefficients of $P_{2}$ polynomial)  in
eqs.(1,3,4)  in two parts in
such a manner that the distributions are positive
defined.
Our expressions for $\Delta q(x) = q^{+}(x)-q^{-}(x)$
($q(x) = q^{+}(x)+q^{-}(x)$) are:
\begin{eqnarray}
\Delta u_{v}(x)&=&x^{-0.462}(1-x)^{3.96}(a_{1}+a_{2}\sqrt
{x}+a_{3}x), \nonumber \\
\Delta
d_{v}(x)&=&x^{-0.670}(1-x)^{4.71}(b_{1}+b_{2}\sqrt{x}+b_{3}x), \nonumber \\
\Delta M(x)&=&x^{-0.800}(1-x)^{9.27}(c_{1}+c_{2}\sqrt{x}), \\
\Delta \delta (x)&=&x^{-0.462}(1-x)^{9.27}d(1+25.0x). \nonumber
\end{eqnarray}

\noindent At this stage we will not take into account gluonic degrees of
freedom.
In the case of $\Delta M$, i.e. total sea polarization, we
assume that there is no
term behaving like $x^{-1.3}$ at small $x$ (we assume that $\Delta M$
and hence all distributions are integrable), which means that
coefficient in this case have to be splitted into equal parts in
$M^{+}$ and $M^{-}$. The next term ($x^{-0.8}$) is relatively
singular in comparison to valence quark distributions. That
means that for $x \rightarrow 0$  the sea contribution
dominates
and hence, proton and neutron spin asymmetries will behave in
the same way in this regime. That is not the trend that
is observed in the existing experimental data. Looking at the
data points we  see that proton asymmetry is positive while neutron negative
for small $x$ values. Nevertheless, we shall consider two kinds
of models: the first (later abbreviated by I) with more singular
($\Delta M \sim x^{-0.8}$) behaviour for $x \rightarrow 0$ and the second
(II) with less singular ($\Delta M \sim x^{-0.3}$). In the second
case the coefficient in
front of $x^{-0.8}$ is also equally divided between $M^{+}(x)$
and $M^{-}(x)$. We would like to stress that such behaviour is
not a Regge type extrapolation in this region.
There are some problems in MRS(A) fit with positivity of quark
sea distributions and the condition
that $q^{+}$ and $q^{-}$ distributions have the same minimal non positive
term determines $d$ in eq.(6) and reduces the number of parameters
from 8 to 7 (or from 9 to 8 when $c_{1} \neq 0$).

To find the unknown parameters in the expressions for polarized
quark distributions we fit our formulas for $A_{1}^{p}$,
$A_{1}^{n}$ and $A_{1}^{d}$ (with eight or seven parameters) to
the experimental data on spin asymmetries, which are
given by:
\begin{eqnarray}
A^{p}_{1}(x)&=&\frac{4\Delta u_{v}(x)+\Delta d_{v}(x)+2.236
\Delta M(x)-3\Delta \delta (x)}{4u_{v}(x)+d_{v}(x)+2.236
M(x)-3\delta (x)}(1+R),  \nonumber \\
A^{n}_{1}(x)&=&\frac{\Delta u_{v}(x)+4\Delta d_{v}(x)+2.236
\Delta M(x)+3\Delta \delta (x)}{u_{v}(x)+4d_{v}(x)+2.236
M(x)+3\delta (x)}(1+R), \\
A^{d}_{1}(x)&=&\frac{5\Delta u_{v}(x)+4.472\Delta M(x)}{5u_{v}(x)+4.472M(x)}
(1-\frac{3}{2}p_{D})(1+R). \nonumber
\end{eqnarray}

The ratio $R=\sigma_{L}/\sigma_{T}$, which vanishes in the Bjorken limit,
is taken
from \cite{whit}, whereas $p_{D}$ is a probability of D-state in deuteron
wave function (equal to $5.8\%$). Spin structure function
$g_{1}^{p}$ is given by:
\begin{equation}
g_{1}^{p}(x,Q^{2})=(4\Delta u_{v}(x)+\Delta d_{v}(x)+2.236
\Delta M(x)-3\Delta \delta (x))/18.
\end{equation}

In this paper we assume that the spin asymmetries do not depend on $Q^{2}$
what is suggested by the experimental data \cite{d,n} and phenomenological
analysis \cite{alt}.
We have made fits using different assumptions about the
sea contributions (I and II) and also for different sets of
experimental data (i.e.: proton+neutron (pn), proton+deuteron
(pd) and proton+neutron+deuteron (pnd)). We do this
because it is  known that neutron
and deuteron spin asymmetries lead to different values of an
integral:
$\Gamma_{1}^{n}(Q^{2})=\int^{1}_{0}g_{1}^{n}(x,Q^{2})\, dx$.
We have
also tried to include gluons along the line of ref.\cite{ross} by
considering effective polarized quark distributions $\Delta
q^{eff}= \Delta q-\frac{\alpha_{s}}{2\pi}\Delta G$ and fitting
the constant in front of gluonic distribution.

The obtained polarized quark distributions $\Delta u(x)$,
$\Delta d(x)$, $\Delta M(x)$ and $\Delta\delta(x)$ can be used to calculate
first
moments. For a given $Q^{2}$ we can write the
relations:
\begin{eqnarray}
\Gamma^{p}_{1}& = &\frac{4}{18}\Delta u+\frac{1}{18}\Delta d+
\frac{1}{18}\Delta s , \nonumber \\
\Gamma^{n}_{1}& = &\frac{1}{18}\Delta u+\frac{4}{18}\Delta d+
\frac{1}{18}\Delta s ,
\end{eqnarray}

\noindent where $\Delta q=\int^{1}_{0}\Delta q(x)\, dx$.

Other combinations (octet and singlet ones) of first moments of
quark polarizations are:
\begin{eqnarray}
a_{3}& = &\Delta u-\Delta d , \nonumber \\
a_{8}& = &\Delta u+\Delta d-2\Delta s , \\
\Delta\Sigma& =&\Delta u+\Delta d+\Delta s , \nonumber
\end{eqnarray}

The results for the integrated quantities (calculated at $4\,{\rm GeV^{2}}$)
after taking into account known QCD corrections (see e.g. Ref.\cite{lar})
could be compared with axial-vector coupling constant
$g_{A}$ and $g_{8}$ known from neutron $\beta$-decay and
hyperon $\beta$-decays (the last one with the help of $SU(3)$
symmetry). We can express
the combination of $\Gamma^{p}_{1}(Q^{2})$ and $\Gamma^{n}_{1}(Q^{2})$:
\begin{equation}
\Gamma^{p}_{1}(Q^{2})-\Gamma^{n}_{1}(Q^{2})=c_{NS}(Q^{2})g_{A}/6,
\end{equation}

\noindent
 where $c_{NS}(Q^{2})$ describes
QCD corrections for non-singlet quantities \cite{lar}
and $g_{A}=1.2573\pm 0.0028$ (see Ref.\cite{cr})
is obtained from the neutron $\beta$-decay.
We get $a_{3}(4\,{\rm GeV^{2}})=
c_{NS}(4\,{\rm GeV^{2}})g_{A}/6=1.11$
and with this value we shall compare $a_{3}$ calculated from our fits.
Another combination of $\Gamma^{p}_{1}(Q^{2})$ and
$\Gamma^{n}_{1}(Q^{2})$ is equal to:
\begin{equation}
\Gamma^{p}_{1}(Q^{2})+\Gamma^{n}_{1}(Q^{2})=5a_{8}(Q^{2})/18+2\Delta
s(Q^{2})/3
\end{equation}

\noindent
with $a_{8}=c_{NS}(Q^{2})g_{8}$ and where $g_{8}=0.58\pm 0.03$
\cite{cr} is obtained from the hyperon $\beta$-decays.
Knowing $c_{NS}(Q^{2})$ we can calculate $a_{8}(4\,{\rm GeV^{2}})=0.515$
and with this number we shall compare the results obtained from our fit.
If we have had very precise experimental data and in the whole $x$ range
there would be
no problems with determination of polarized quark distributions. Unfortunately
that is not the case yet. Actually, from the experiment we have
information on $\Gamma^{p}_{1}$ and $\Gamma^{n}_{1}$.
The combination $\Gamma^{p}_{1}$-$\Gamma^{n}_{1}$ is directly
connected to $g_{A}$ experimental
quantity modified by QCD corrections.  On
the other hand $\Gamma^{p}_{1}$+$\Gamma^{n}_{1}$ is the
combination of $a_{8}$ and $\Delta s$ and it comes out that
the fits are not sensitive enough to determine $a_{8}$ and $\Delta s$
separately in a stable way. The value of $a_{8}$ and $\Delta s$
are different for models I and II and
different subsets of data. To stabilize the determination of
parameters we assume in addition that $a_{8}=0.515$ with
$0.1$ as artificial theoretical error. We will not present the results of
all our fits. Some examples are given in the Table 1.

\begin{center}
{\bf Table 1}
\end{center}
 The first moments of polarized distributions (see
eqs.(9) and (10)). The strange sea polarization $\Delta s$ is
connected to the total sea polarization by the relation:
$\Delta s=0.196 \Delta M$. We present figures for two type of
models: I and II. We have made our fits taking different spin asymmetries,
namely: for  proton (p) (P stands for all proton data with the inclusion
of newest SLAC E143 points), neutron (n) and deuteron (d) target.
\begin{center}
\begin{tabular}{|c||r|r|r|r||r|} \hline
  &I(p,n,d)&II(p,n,d)&II(p,d)&II(p,n)&II(P,n) \\ \hline\hline
$\Gamma^{p}_{1}$&0.156&0.146&0.145&0.148&0.144 \\ \hline
$\Gamma^{n}_{1}$&-0.045&-0.055&-0.076&-0.038&-0.042\\ \hline
$g_{A}$&1.205&1.201&1.330&1.114&1.112 \\ \hline
$g_{8}$&0.508&0.512&0.515&0.523&0.478 \\ \hline
$\Delta \Sigma $&0.384&0.297&0.204&0.373&0.348 \\ \hline
$\Delta M$&-0.210&-0.366&-0.529&-0.255&-0.221 \\ \hline
\end{tabular} \\
\end{center}

It would
be natural to use all the existing data for the determination of
polarized quark distributions. We see that in this case fits I
and II lead to similar results but formally fit I
has one parameter more and higher $\chi^{2}$ per degree of
freedom. From the Table 1 we see how integrated quantities:
$\Gamma^{p}_{1}$, $\Gamma^{n}_{1}$, $a_{3}$, $a_{8}$,
$\Delta\Sigma$ and $\Delta M$ depend on usage of  different subsets
of data. It seems that the closest value to the expected
theoretical value for $a_{3}=1.11$ we get when we use the
combination of all proton and neutron SLAC data. In the ratio
$a_{8}/a_{3}$ QCD corrections cancel so it can be expressed
directly through the quantities known from low energy decay
experiments i.e. $g_{8}/g_{A}=0.46 \pm 0.02$. For the combination of proton and
neutron SLAC data the ratio calculated from the fit ($a_{8}/a_{3}=0.47$) is
closest to the value expected from experiment. We get for this
fit $\chi^{2}=20.47$. If we add to this value the figure 6.93 corresponding
to deuteron contribution (from the SMC experiment) we get the value 27.4 what
is
very close to the the $\chi^{2}$ value obtained from the fit to
all ( proton, neutron and deuteron) data, namely 27.1. If one
starts, on the other hand, with the fit to proton and deuteron data and
adds to the $\chi^{2}$ value the number corresponding to
neutron contribution one
gets $\chi^{2}=37.1$. It seems that it is possible to get a
satisfactory deuteron asymmetry using proton+neutron data and
the result is not as good when we start with the combination
of proton and deuteron data (in this case $a_{3}=1.33$, the value which
is far too big). So we have decided to use combination of proton and
neutron ($^{3}He$) spin asymmetry data to calculate the
parameters of the polarized quark distributions. We present in
Figs.(1a,1b,2 and 3) the comparison of our fit II(p,n) with the
experimental asymmetries for proton (1a,1b) neutron (2) and
deuterium (3) target. In the last figure our curve should be treated
as a prediction, because we do not take deuteron data into account
in this fit. The parameters given in the first row of the Table 2
correspond to the considered model, namely the one from the
fifth column  of  Table 1.

\begin{center}
{\bf Table 2} \\
\end{center}
The coefficients of polarized distributions (see eqs.(6))
for type II models. The second row corresponds to the figures
gotten in a fit with the inclusion of the new proton data from
the E143 experiment at SLAC.
\begin{center}
\begin{tabular}{|c||c|c|c|c|c|c|c|c|} \hline
  &$a_{1}$&$a_{2}$&$a_{3}$&$b_{1}$&$b_{2}$&$b_{3}$&$c_{2}$&$d$\\ \hline\hline
II(p,n)&0.924&-3.237&11.30&-0.029&-0.163&-1.644&-0.993&-0.015 \\ \hline
II(P,n)&0.929&-3.005&10.24&-0.066&-0.064&-1.644&-0.861&-0.013\\ \hline
\end{tabular} \\
\end{center}

The obtained quark distributions lead
to the following integrated quantities:
$\Delta u=0.91$ ($\Delta u_{v}=1.04$), $\Delta d=-0.33$
($\Delta d_{v}=-0.18$) and $\Delta s=-0.07$.
It gives the amount of sea polarization $\Delta M=-0.26$.
We see that our integrated quantities ($\Gamma^{p}_{1}$ and $\Gamma^{n}_{1}$)
differ slightly from the
values quoted by the experimental groups. The experimental figures are
calculated directly from
the experimental points with the assumption of Regge type behaviour at small
$x$. On the other hand our polarized quark distributions satisfy all
the constraints taken implicitly into account in fits to
unpolarized data.

We have also tried to consider gluonic contributions which were proposed
as a solution of spin "crisis" . We have taken into account gluons
by considering effective polarized quark distributions $\Delta
q^{eff}=\Delta q-\frac{\alpha_{s}}{2\pi}\Delta G$ where
\begin{equation}
\Delta G(x)=fx^{-0.3}(1-x)^{5.3},
\end{equation}

\noindent with a new $f$ constant which was fitted.
In the basic fit (II(p,n)) we have got $\alpha_{s}\Delta G/2\pi=-0.08$ (
which corresponds to $\Delta G=-1.84$ for $\alpha_{s}=0.28$).
This means that
the sign of the gluonic contribution is opposite to what is
expected theoretically and effective $\Delta \Sigma^{eff}$ is
bigger than $\Delta \Sigma$ coming from quark distributions.
We conclude that from the point of view of fitting we do
not need gluonic contribution.

The Fig.(4) shows the comparison of $g^{p}_{1}(x)$ calculated
from our fit with experimental points. We observe the substantial
growth of $g^{p}_{1}$ for small $x$ values in our model.

The last column of Table 1 corresponds to the inclusion of
recent data for spin asymmetries from the experiment E143 at
SLAC. We see that the inclusion of new data changes integrated
quantities only slightly what we consider as a positive fact.
In the case of the good fit the additional experimental data
should not change parameters in a drastic way. The
parameters corresponding to this fit are given in the second row
of Table 2 whereas the fitted curve is presented in Fig.(5)
together with the one for fit II(p,n).
We also would like to point out that with the new
data from E143 included the fit of the type I model coincides with that
of type II, which means that the more singular behaviour of the sea
is eliminated by fitting the formulas to the experimental data. However,
one should be aware of the fact that in the last fit we have much more
proton than neutron data points.

Now, we would like to make some comments about the $x \rightarrow 1$
behaviour of valence quark distributions. Looking at the data points for
proton spin asymmetry the value close to 1 at $x \sim 1$ is preferred,
whereas for neutron and deuteron case values close to 0 seem to be
natural (such observation is fragile due to the big experimental errors
in this $x$ region). In our approach we can give predictions for the
behaviour of polarized quark distributions and spin asymmetries in the
$x \rightarrow 1$ limit. In our fit we get
$A^{p}_{1}=A^{n}_{1}= A^{d}_{1}=0.89$.

Starting from the new, improved version of the MRS fit \cite{mrs1}
to the unpolarized deep inelastic data we have made a fit to
proton, neutron ($^{3}He$) and deuteron spin asymmetries in
order to obtain polarized quark parton distributions. We have
discussed two kinds of models with more and less singular sea
contribution at small $x$ (both models being a possible consequence of the
unpolarized distributions) and different combination of proton, neutron
and deuteron spin asymmetries data. To stabilize the fits we add
the experimental information on octet quantity $a_{8}$. We have
calculated the parameters of the polarized quark distributions
from the combination of proton and neutron spin
asymmetries data (without the SMC deuteron data). The inclusion
of the new proton SLAC data from E143 experiment modifies our fit
only slightly, which is a positive feature. We do not
need gluonic contributions to be taken into account, i.e. the fit
with gluons is worse.
The new consistent data for spin asymmetries (mainly for deuteron and
neutron) could help us to determine polarized quark distributions
and to resolve the doubts about the $x\rightarrow 0$ and
$x\rightarrow 1$ behaviour of such functions.

{\bf Figure captions}

\begin{itemize}
\item[ Figure 1a] The comparison of spin asymmetry on protons
(data points are from SLAC (E80, E130) and CERN (EMC,SMC) experiments)
with the curve gotten from our fit II(p,n) (eqs.(6,7) and the first row
of Table 2).
\item[ Figure 1b] The same as in figure 1a but with $x$ in logarithmic scale.
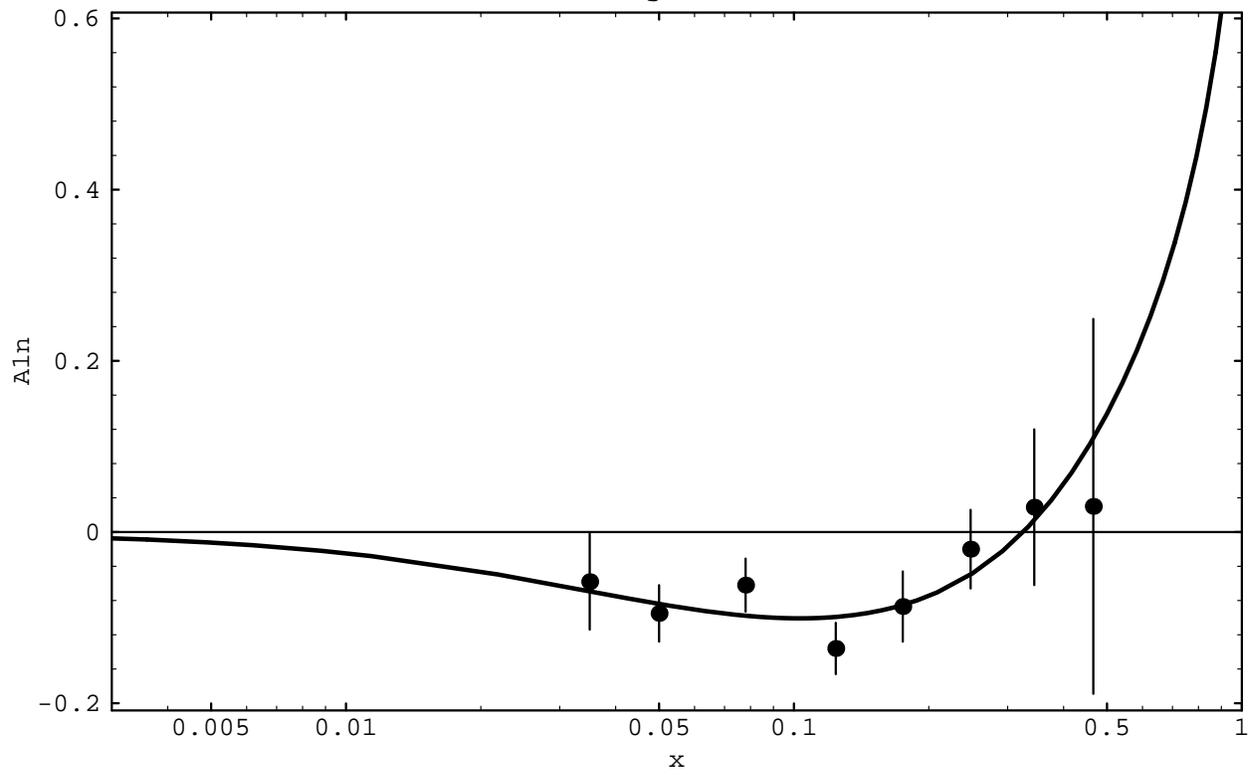
\item[Figure 2 \ ] The comparison of spin asymmetry on neutrons
(SLAC E142 data) with the curve gotten from our fit.
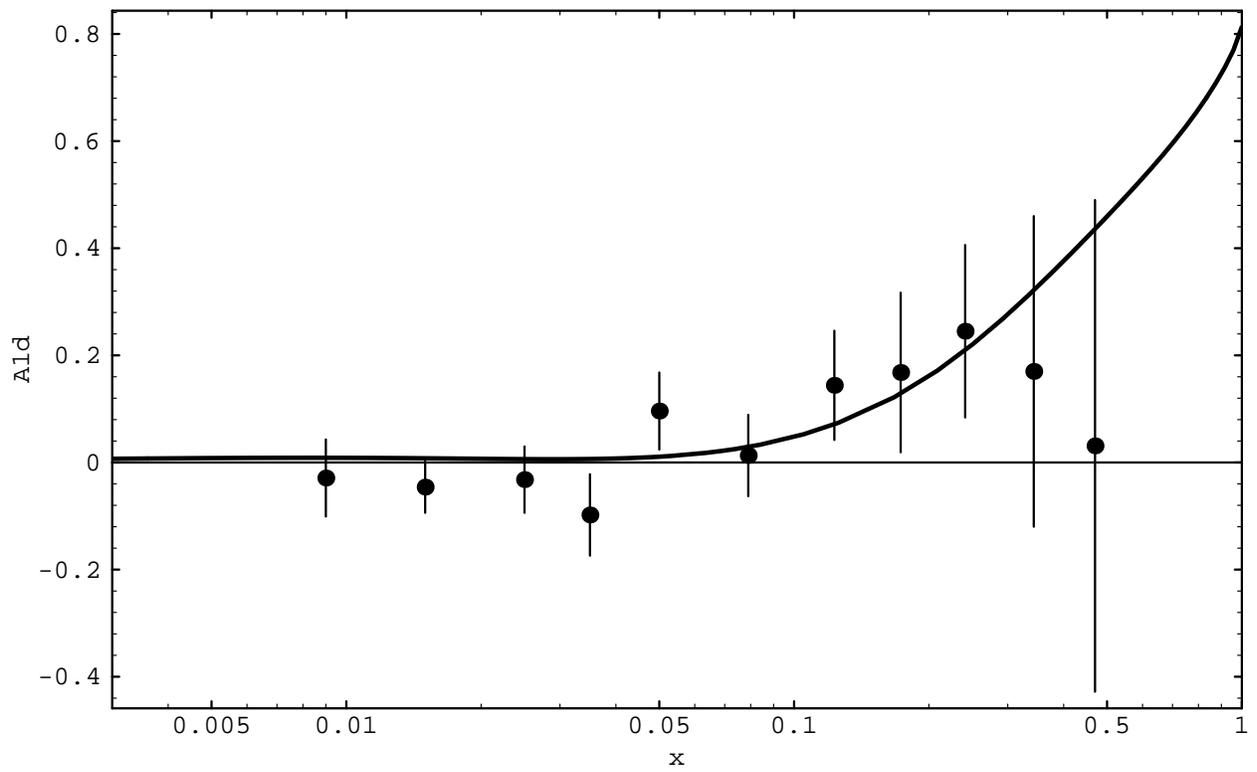
\item[Figure 3 \ ] Our prediction for deuteron asymmetry compared
with the SMC data.
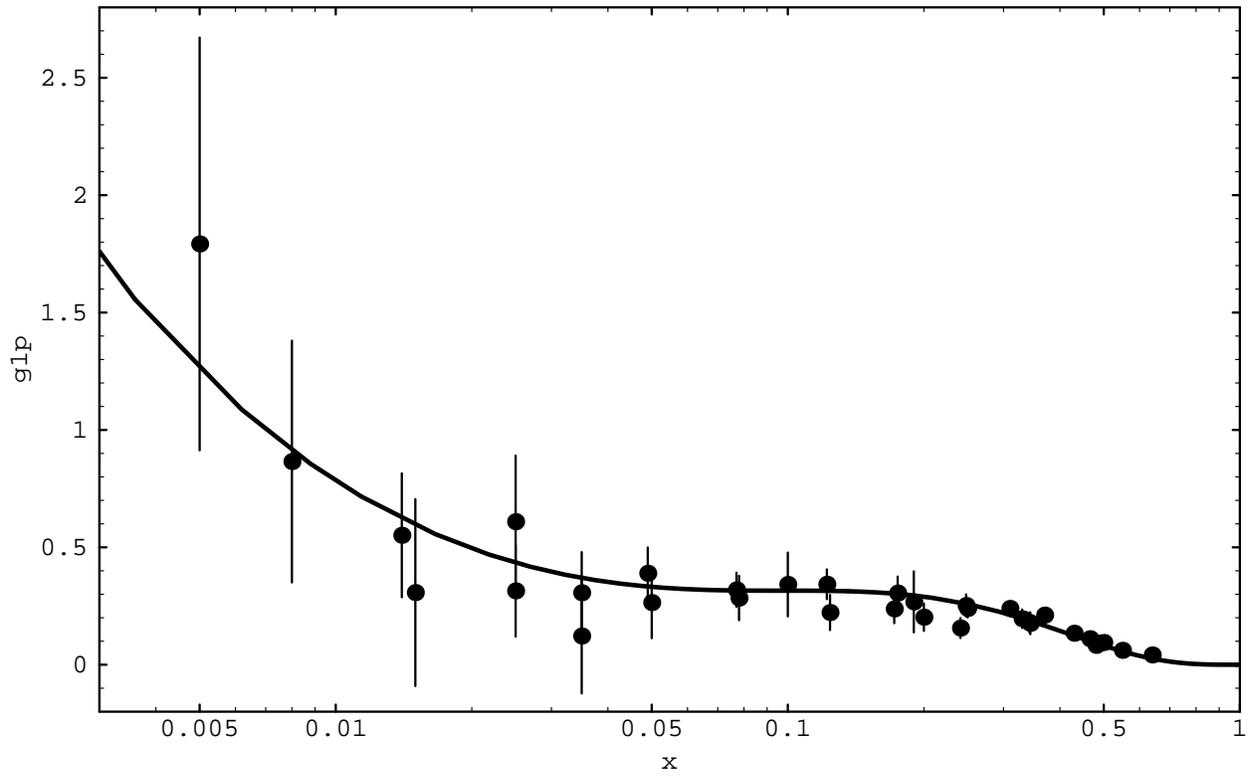
\item[Figure 4 \ ] The data for $g_{1}^{p}(x)$ structure function
with the curve gotten using the parameters of fit II(p,n).
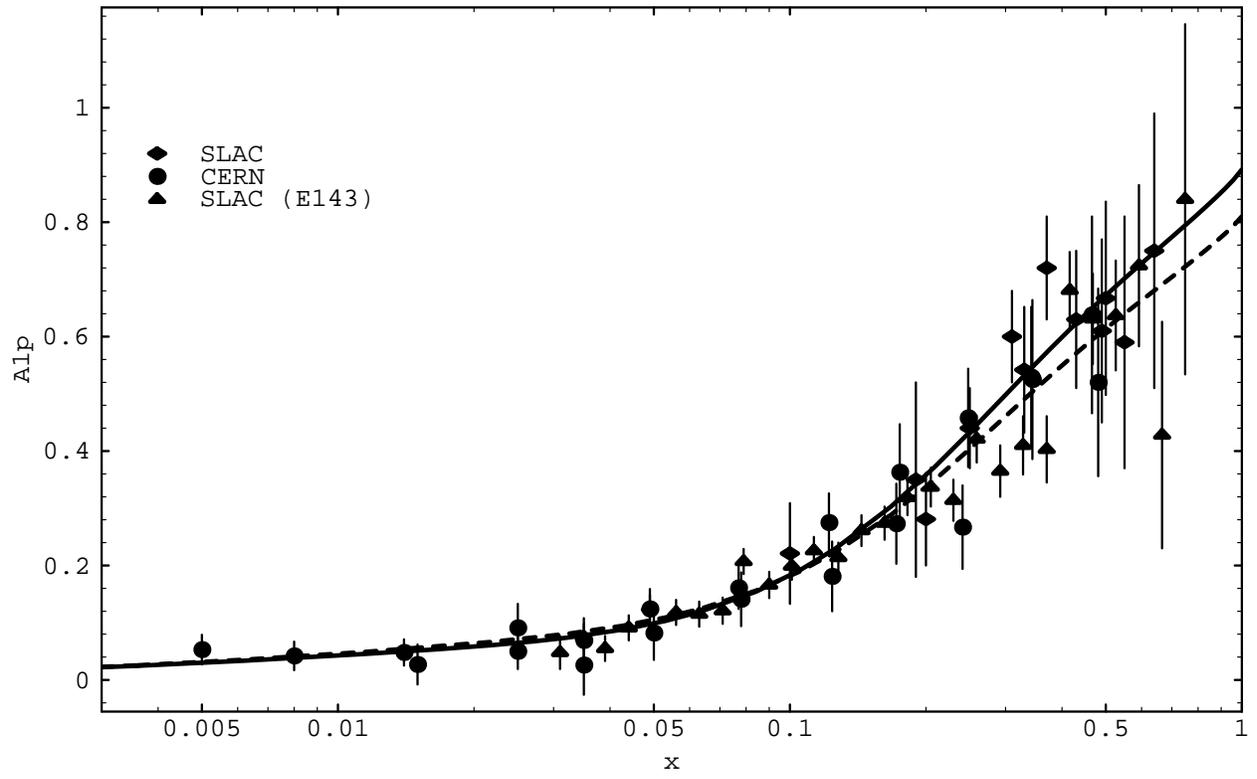
\item[Figure 5 \ ] The same as in Fig.(1b) but with the new
experimental points (E143) from SLAC included.
The dashed curve corresponds to the fit II(P,n), which takes into
account the E143 experiment data.
\end{itemize}

\end{document}